\documentclass[times]{weauth2}
\usepackage{moreverb}
\usepackage{subfigure}
\usepackage{graphics}
\usepackage{graphicx}
\usepackage{epsfig}
\usepackage{epstopdf}
\usepackage{booktabs}
\usepackage{amsmath}
\usepackage[colorlinks,bookmarksopen,bookmarksnumbered,citecolor=red,urlcolor=red]{hyperref}

\graphicspath{{./fig/}} 			

\newcommand\BibTeX{{\rmfamily B\kern-.05em \textsc{i\kern-.025em b}\kern-.08em
T\kern-.1667em\lower.7ex\hbox{E}\kern-.125emX}}

\def\volumeyear{2014}

\begin{document}

\runningheads{Sakagami Y. \emph{et al.}}{Logarithmic Wind Profile: A Stability Wind Shear Term}

\articletype{RESEARCH ARTICLE}

\title{Logarithmic Wind Profile: A Stability Wind Shear Term}

\author{Yoshiaki Sakagami\affil{1,2}, Pedro A. A. Santos\affil{2}, Reinaldo Haas\affil{3}, Julio C. Passos\affil{2},  Frederico F. Taves\affil{4}}

\address{\affilnum{1} Dept. of Health and Service, Federal Institute of Santa Catarina, Florian{\'{o}}polis, Brazil
\newline \affilnum{2} Dept. of Mechanical Engineering, Federal University of Santa Catarina, Florian{\'{o}}polis, Brazil
\newline \affilnum{3} Dept. of Physics, Federal University of Santa Catarina, Florian{\'{o}}polis, Brazil
\newline \affilnum{4} Tractebel Energia S.A. (GDF Suez), Florian{\'{o}}polis, Brazil  }

\corraddr{ Yoshiaki Sakagami,Dep. of Health and Service, Federal Institute of Santa Catarina, Florian{\'{o}}polis, Brazil.
\newline E-mail: yoshi@ifsc.edu.br }

\begin{abstract}
A stability wind shear term of logarithmic wind profile based on the terms of turbulent kinetic energy equation is proposed. The fraction influenced by thermal stratification is considered in the shear production term. This thermally affected shear is compared with buoyant term resulting in a stability wind shear term. It is also considered Reynolds stress as a sum of two components associated with wind shear from mechanical and thermal stratification process. The stability wind shear is responsible to Reynolds stress of thermal stratification term, and also to Reynolds stress of mechanical term at no neutral condition. The wind profile and its derivative are validated with data from Pedra do Sal experiment in a flat terrain and 300m from shoreline located in northeast coast of Brazil. It is close to the Equator line, so the meteorological condition are strongly influenced by trade winds and sea breeze. The site has one 100m tower with five instrumented levels, one 3D sonic anemometer, and a medium-range wind lidar profiler up 500~m. The dataset are processed and filter from September to November of 2013 which results in about 550 hours of data available. The results show the derivative of wind profile with $R^2$ of 0.87 and RMSE of 0.08~$m~s^{-1}$. The calculated wind profile performances well up to 400~m at unstable condition and up to 280~m at stable condition with $R^2$ better than 0.89. The proposed equation is valid for this specific site and is limited to a stead state condition with constant turbulent fluxes in the surface layer.
\end{abstract}


\maketitle

\section{Introduction}

The study of wind speed profile in the atmospheric boundary layer has progressed recently by new technologies such as sodar and lidar profilers, and better eddy covariance systems. The theory of wind profile that is based on semi-empirical the logarithmic law that is well correlated at neutral conditions. However, the atmosphere is rarely neutral, and turbulence has a strong influence on the wind speed profile. The turbulence is related to atmospheric stability concept when unstable flow become or remain turbulent and stable flow become or remain laminar \cite{stull1988}. The cornerstone of this problem is that the wind profile need to be correct not just by the slope of logarithmic profile but its shape \cite{Tennekes1973}. It also depends on a stability parameter that is not well understood because of stochastic nature of turbulent process. The most common parameter used to quantify the atmospheric stability is the Obukhov length (L) based on the Monin-Obukhov similarity theory \cite{obukhov1954}. Many experiments conducted in the last 50 years have showed a good agreements to the similarity theory with an accuracy of about 10-20\% \cite{kaimal1990}, \cite{hess1980}, \cite{hogstron1988}, \cite{foken2006}. They suggests many different parameters to the universal similarity function and the accepted function presently is the re-formulated function from Businger et al. \cite{Businger1971} proposed by H{\"{o}}gstr{\"{o}}m \cite{hogstron1988} \cite{foken2006}. However, the experiments are located at specific sites and limited to particular meteorological conditions. Recent studies about atmospheric stability related to wind turbines performance diverges for each specific site condition \cite{peinke2007}, \cite{wagner2009},\cite{rareshide2009}, \cite{wharton2011},\cite{pena2008}), \cite{wharton2012}. This is one of the gaps for wind energy applications and a better understanding of the wind profile and other properties such as turbulence and stability is necessary \cite{Banta2013}. Other important studies are related to the improvements of micro siting models for wind energy application \cite{lange2006} and the numerical weather predictions models \cite{pena2011} as the boundary layer in those models are still poor represented \cite{floors2013}. The limits of the actual theory and particular results suggest that a local experiment is necessary when wind profile need to be accurate predict at a specific site. Thus, an experiment is set in Pedra do Sal Wind Farm located in north-east coast of Brazil when the meteorological condition are strongly influenced by trade winds and sea breeze. A instrumented tower of 100m with five levels of wind measurements, one sonic 3D anemometer and one lidar wind profiler is used. Based on the analysis of this data and the turbulent kinetic energy equation, a stability wind shear term of the logarithmic wind profile is proposed. The equation is compared with data of this experiment and is showed in the results section. The limits of this equation and conclusions are presented in the last section.

\section{Theory}

\subsection{Shear Components}

The hypothesis of Eddy viscosity, suggested by J. Boussinesq in 1877 \cite{arya2001}, assumes that Reynolds stress ($\tau$) can be analogous with Newton's law of molecular viscosity. Considering the coordinate system aligned with the mean flow and horizontal mean gradients neglected in comparison to those in vertical at the surface layer, $\tau$ can be expressed as:

\begin{equation} \label{eq:eddy}
\tau=-\rho \overline{u^{'}w^{'}}=\rho K_m \frac{\partial{\bar{U}}}{\partial{z}}
\end{equation}

where the $K_m$ is known as eddy exchange of momentum or simply eddy viscosity, $\rho$ is the air density, $\bar{U}$ is the horizontal mean flow and $u$ and $w$ are the horizontal and vertical wind speed component respectively and the apostrophes are the deviations from their means. Eq. \ref{eq:eddy} represents the total Reynolds stress, but it maybe be view as a sum of two components assuming that at a given height forced and free convection are mutually independent \cite{Sellers1962}. One part has a eddy viscosity coefficient associated with forced convection and another due free convection. Similar to this assumption, it is assume that Reynolds stress can be divide in two components as well, but each component is related to wind shear instead of eddy viscosity:

\begin{equation} \label{eq:tau}
\tau=\tau_m+\tau_s
\end{equation}

where $\tau_m$ term is associated with wind shear produced/loss by mechanical process ($\partial{\bar{U}_m}/\partial{z}$) and $\tau_s$ term is associated with wind shear produced/consumed by thermal stratification process ($\partial{\bar{U}_s}/\partial{z}$). Then, combining Eq. \ref{eq:eddy} and Eq. \ref{eq:tau}:

\begin{equation}
\rho K_m \frac{\partial{\bar{U}}}{\partial{z}}= \rho K_m \frac{\partial{\bar{U}_m}}{\partial{z}}+ \rho K_m \frac{\partial{\bar{U}_s}}{\partial{z}}
\end{equation}

It is considered the flow incompressible and coefficients of eddy viscosity at neutral condition in the surface layer. Then, the sum of Reynolds stress can be express as a simply sum of wind shear:

\begin{equation} \label{eq:shear_comp}
\frac{\partial{\bar{U}}}{\partial{z}}= \frac{\partial{\bar{U}_m}}{\partial{z}}+\frac{\partial\bar{{U}_s}}{\partial{z}}
\end{equation}

The mechanical and thermal stratified process associated with wind shear are described in details in next section.

\subsection{Thermal Stratified Shear}

The turbulence kinetic energy (TKE) is one of the most important variable in the boundary layer and its budget equation describes the physical process that produce/consume the turbulence \cite{stull1988}. The simplify form of TKE budget equation, based on horizontal homogeneity, neglect subsidence and coordinate system aligned with the mean flow can be expressed by:

\begin{equation} \label{eq:tke}
\frac{\partial{\bar{e}}}{\partial{t}}=
\\\frac{g}{\theta_v}(\overline{w^{'}\theta_v^{'}})
\\ -\overline{u^{'}w^{'}}\frac{\partial{\bar{U}}}{\partial{z}}
\\-\frac{\partial{(\overline{w^{'}e)}}}{\partial{z}}
\\-\frac{1}{\rho}\frac{\partial{(\overline{w^{'}p^{'})}}}{\partial{z}}-\varepsilon
\end{equation}

where the terms represent: the storage, buoyant, shear, transport, pressure perturbation and dissipation, respectively. The turbulence kinetic energy is $e$, $\theta_v$ is the potential virtual temperature, $g$ is the gravitational acceleration,$u$ and $w$ are the horizontal and vertical wind speed component respectively, $\bar{U}$ is the horizontal mean flow, $p$ is the atmospheric pressure and $\rho$ is the air density. The variables with apostrophes are the deviations from their means and the bar symbol is the variable average. The difficult to study the turbulence is that it a dissipative process and its equation is not conservative \cite{stull1988}. Then, the atmospheric stability is usually analysed by comparing terms of this equation as dimensionless ratio such as flux Richardson number ($R_{f}$), even though the other terms in the TKE budget are also important:

\begin{equation}
R_{f}=\frac{\frac{g}{\theta_v}(\overline{w^{'}\theta_v^{'}})}{-\overline{u^{'}w^{'}}\frac{\partial{\bar{U}}}{\partial{z}}}
\end{equation}

This ratio is modified by a new Richardson number, named as $Ri_{s}$. It is assumed that $Ri_{s}$ represents only the wind shear associated with thermal stratification ($\partial{\bar{U}_s}/\partial{z}$).

\begin{equation} \label{eq:Ris}
Ri_{s}=\frac{\frac{g}{\theta_v}(\overline{w^{'}\theta_v^{'}})}{-\overline{u^{'}w^{'}}\frac{\partial{\bar{U}_{s}}}{\partial{z}}}
\end{equation}

It means that $Ri_{s}$ is proportion of the rate at which the energy is produced/consumed by the buoyant term and the rate at which the energy is produced/loss by the fraction of wind shear related to thermal stratification. This proportion is assumed to be constant in steady-state condition, so $Ri_{s}$ is considered a constant coefficient in the equation and can be determined from experimental data (Fig. \ref{fig:shear}).

\begin{equation}
-\overline{u^{'}w^{'}}\frac{\partial{\bar{U}_s}}{\partial{z}}Ri_{s}=\frac{g}{\theta_v}(\overline{w^{'}\theta_v^{'}})
\end{equation}

The turbulent fluxes of momentum and heat can also be rewritten in terms of sensible heat flux ($H_{s}$) and friction velocity ($u_{*}$) \cite{arya2001}. Then, the wind shear related to thermal stratification can be express as:

\begin{equation} \label{eq:shear_thermal}
\frac{\partial{\bar{U}_s}}{\partial{z}}=\frac{-g H_s}{\rho c_p \theta_v u_{*}^2 Ri_{s}}
\end{equation}

where $c_p$ is the specific heat of dry air at constant pressure. This term is simply named as stability wind shear. It is direct responsible to Reynolds stress $\tau_s$ and also to influence the mechanical wind shear at no neutral condition, which is show in the next section.

\subsection{Mechanical Shear}

The Reynolds stress $\tau_m$ is associated with wind shear produced by mechanical process ($\partial{\bar{U}_m}/\partial{z}$). At neutral condition, the mechanical wind shear is $u_{*}/(\kappa z)$, where $\kappa$ is the von K\'{a}rm\'{a}n constant and z is the height. However, the mechanical wind shear is influenced by thermal stratification process at no neutral conditions. Then, it is assume that mechanical Reynolds stress can be divided in two terms:

\begin{equation}
\tau_m=\tau_n+\tau_{nn}
\end{equation}

where $\tau_n$ is associated with wind shear at neutral condition ($\partial{\bar{U}_n}/\partial{z}$) and $\tau_{nn}$ is associated with the fraction of wind shear influenced by thermal stratification process at no neutral condition ($\partial{\bar{U}_{nn}}/\partial{z}$).

\begin{equation}
\rho K_m \frac{\partial{\bar{U}_m}}{\partial{z}}= \rho K_n \frac{\partial{\bar{U}_n}}{\partial{z}}+ \rho K_{nn} \frac{\partial{\bar{U}_{nn}}}{\partial{z}}
\end{equation}

The eddy viscosity can not be considered at neutral condition to all coefficients in this case. Each term has its own coefficient of eddy viscosity related to its wind shear. Then,  $K_m=l_m^2|\partial{\bar{U}_m}/\partial{z}|$, $K_n=l_m^2|\partial{\bar{U}_n}/\partial{z}|$ and $K_{nn}=l_m^2|\partial{\bar{U}_{nn}}/\partial{z}|$, and $l_m$ is the mean mix length.

\begin{equation}
\rho l_m^2 \Bigg(\frac{\partial{\bar{U}_m}}{\partial{z}}\Bigg)^2= \rho l_m^2 \Bigg(\frac{\partial{\bar{U}_n}}{\partial{z}}\Bigg)^2+ \rho l_m^2 \Bigg(\frac{\partial{\bar{U}_{nn}}}{\partial{z}}\Bigg)^2
\end{equation}

It is considered the flow incompressible and the mixing length constant at each height in the surface layer ($l_m=\kappa z$).

\begin{equation} \label{eq:shear_neutro}
\Bigg(\frac{\partial{\bar{U}_m}}{\partial{z}}\Bigg)^2= \Bigg(\frac{\partial{\bar{U}_n}}{\partial{z}}\Bigg)^2+ \Bigg(\frac{\partial{\bar{U}_{nn}}}{\partial{z}}\Bigg)^2
\end{equation}

Simplifying the Eq. \ref{eq:shear_neutro}, the mechanical wind shear influenced by thermal stratification process at no neutral condition can be express as:

\begin{equation} \label{eq:shear_mechanical}
\frac{\partial{\bar{U}_m}}{\partial{z}}= \sqrt{\Bigg(\frac{\partial{\bar{U}_n}}{\partial{z}}\Bigg)^2+ \Bigg(\frac{\partial{\bar{U}_{nn}}}{\partial{z}}\Bigg)^2}
\end{equation}

where the wind shear at neutral condition is:

\begin{equation}
\frac{\partial{\bar{U}_n}}{\partial{z}}=\frac{u_{*}}{\kappa z}
\end{equation}

and fraction of wind shear affected by thermal process at no neutral condition is assume to be equals to stability wind shear (Eq. \ref{eq:shear_thermal}).

\begin{equation}
\frac{\partial{\bar{U}_{nn}}}{\partial{z}}=\frac{\partial{\bar{U}_s}}{\partial{z}}
\end{equation}

\subsection{Total Shear}

The total wind shear can be express by combining Eq.~\ref{eq:shear_comp}, Eq.~\ref{eq:shear_thermal}, Eq.~\ref{eq:shear_mechanical}.

\begin{equation} \label{eq:shear_total}
\frac{\partial{\bar{U}}}{\partial{z}}= \sqrt{\Bigg(\frac{u_{*}}{\kappa z}\Bigg)^2 + \\
\Bigg(\frac{-g H_s}{\rho c_p \theta_v u_{*}^2 Ri_{s}}\Bigg)^2} \\
-\frac{g H_s}{\rho c_p \theta_v u_{*}^2 Ri_{s}}
\end{equation}

\subsection{Wind Profile}

The terms of the Eq.~\ref{eq:shear_total} can be renamed to simplify the equation.

\begin{equation}\label{eq:shear}
\frac{\partial{\bar{U}}}{\partial{z}}=\psi+\psi_s
\end{equation}

where

\begin{equation}
\psi=\sqrt{\Bigg(\frac{u_{*}}{\kappa z}\Bigg)^2+\psi_s^2}
\end{equation}

and

\begin{equation}
\psi_s=\frac{-g H_s }{\rho c_p \theta_v u_{*}^2 Ri_{s}}
\end{equation}

The turbulent fluxes of heat and momentum can be considered constants with the height in the surface layer \cite{obukhov1954}, and the wind profile can be found by integrating Eq.~\ref{eq:shear}.

\begin{equation}
\int{}\partial{\bar{U}}=\int{(\psi+\psi_s)}\partial{z}
\end{equation}

Solving this integral, the logarithmic wind speed profile with wind shear stability term can be written as:

\begin{equation}\label{eq:prof}
\bar{U}(z)=\frac{u_{*}}{\kappa} \ln{\Bigg(\frac{z}{z_o}\Bigg)}\\
+(\psi z-\psi_{o} z_o)- \frac{u_{*}}{\kappa} \ln{\Bigg(\frac{\psi z + \frac{u_{*}}{\kappa}}{\psi_{o} z_o+ \frac{u_{*}}{\kappa}}\Bigg)}+\psi_s z
\end{equation}

where $z_o$ is the coefficient roughness and $\psi_{o}$ is the mechanical shear at $z_o$.

\subsection{Reference Shear}

The dimensionless shear ($\phi_{m}$) introduced by Obukhov \cite{obukhov1954} is widely used as a parameter to study the influence of the atmospheric stability on the wind profile.

\begin{equation}
\phi_{m}=\frac{\kappa z}{u_{*}}\frac{\partial{\bar{U}(z)}}{\partial{z}}
\end{equation}

The ratio of von K\'{a}rm\'{a}n constant and friction velocity ($\kappa/u_{*}$) in the equation can be viewed as a slope correction due mechanical stability. Therefore, the dimensionless shear is not in fact the slope of the wind profile observed, but a slope corrected by friction velocity. In order to analyse the slope of wind profile without mechanical corrections, the ratio $\kappa/u_{*}$ is omitted from dimensionless shear equation, and a simplified reference shear ($\phi_{r}$) is adopted.

\begin{equation} \label{eq:refshearR}
\phi_{r}=\frac{z\partial{\bar{U}(z)}}{\partial{z}}
\end{equation}

The reference shear can be calculated using the measurements of the wind speed at several heights and approximating the profile by a second-order polynomial \cite{hogstron1988}.

\begin{equation}
u(z)=u_{o}+Aln(z)+B(ln(z))^{2}
\end{equation}

where the coefficients $u_{o}$, A and B can be found by adjusting the best polynomial fitting.

The polynomial equation above is derived, so the reference shear observed can be expressed as:

\begin{equation}
\phi_{r}= A+2Bln(z_m)
\end{equation}

where $z_m$ is the geometric mean height for logarithmic approximation \cite{arya2001}.

\begin{equation} \label{eq:zp}
z_m=\sqrt{z_{high}z_{low}}
\end{equation}

where, high and low labels are the top and button levels of the measurements. The theoretical reference shear ($\phi_{s}$) is estimated by Eq.~\ref{eq:shear} multiplied by $z_m$.

\begin{equation}\label{eq:refshearS}
\phi_{s}= z_m(\psi+\psi_s)
\end{equation}

The observed and calculated reference shear are compared by linear fitting and $Ri_{s}$ can be estimated at the best coefficient of determination.

\section{Experiment}

\subsection{Site description}

\begin{figure}[ht]
\centering
\includegraphics[scale=.35]{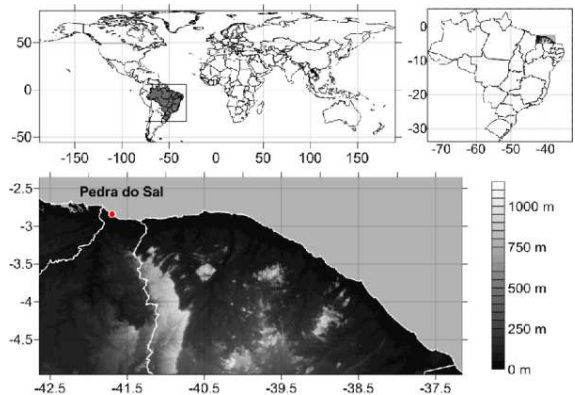}
\caption{Pedra do Sal experiment location at northeast Brazilian coast indicated with a red dot on the map. A concave coastline and topograph surrounding determine the local characteristics. The mountain range is about 80~km from the site and still influences the trade winds from southeast.}
\label{fig:map}
\end{figure}

\begin{figure}[ht]
\centerline{%
\begin{tabular}{c@{\hspace{0.01pc}}c}
\includegraphics[width=7cm]{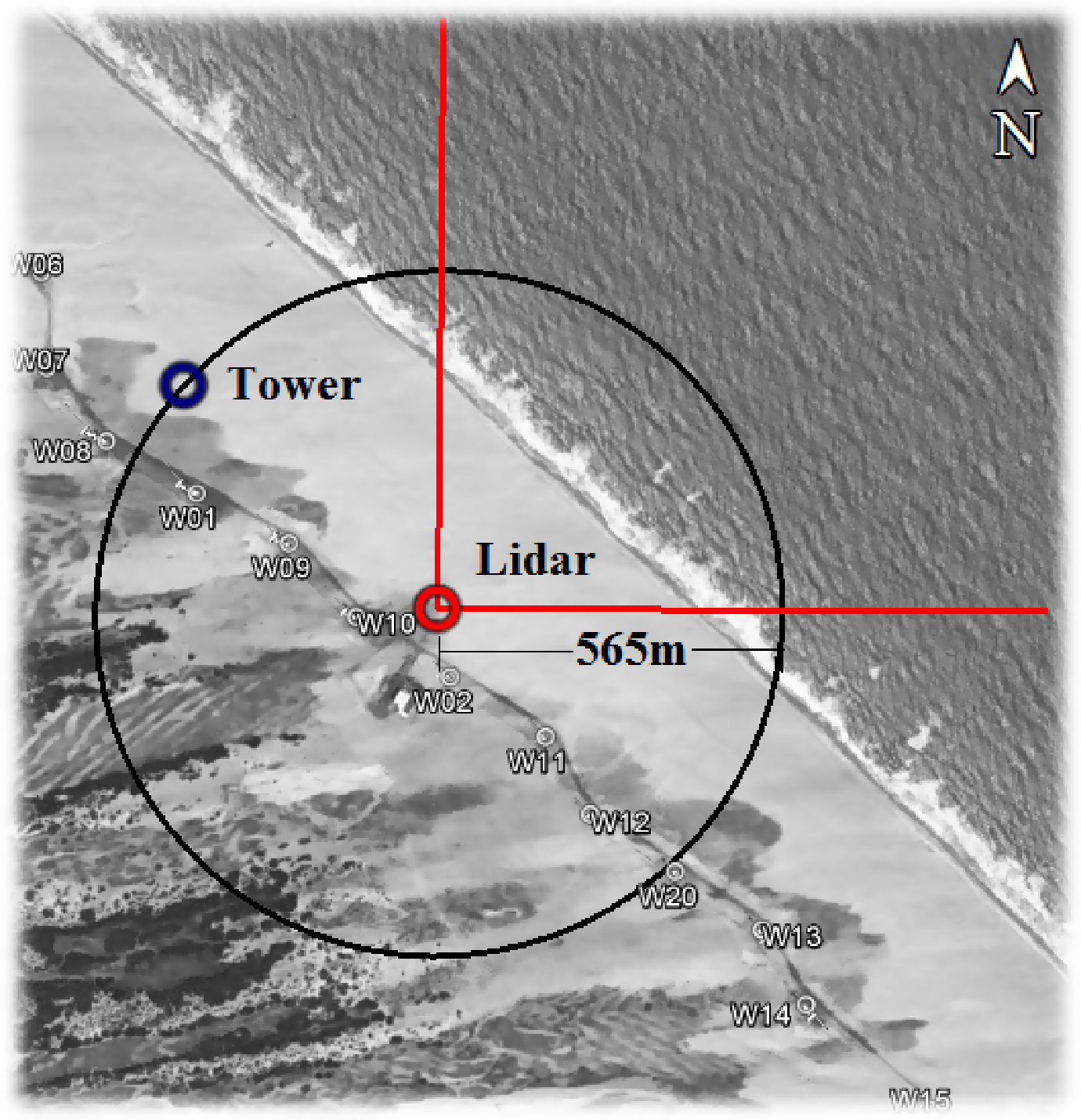} &
\includegraphics[width=8.2cm]{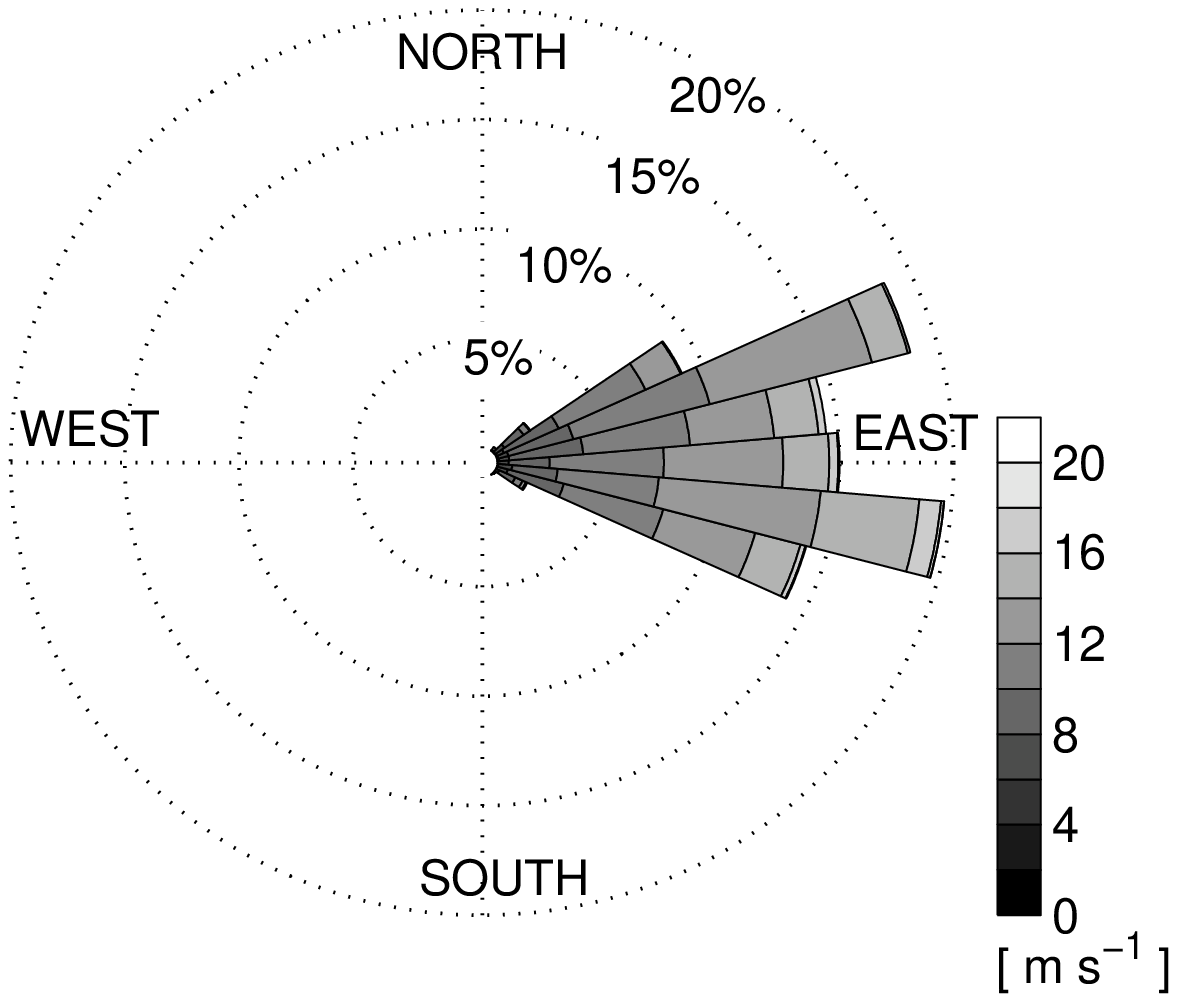} \\
a.~~Sector $0^{\circ}$ to $90^{\circ}$   & b.~~ Wind Rose
\end{tabular}}
\caption{(a.) Satellite image from Pedra do Sal Wind Farm. The picture shows wind turbines aligned with the coast, the position of the lidar marked in red circle and the meteorological tower in blue circle. The black circle indicates the distance from lidar to tower, and the red lines represents the limits of the sector free of wake  (0$\,^{\circ}$ to 90$\,^{\circ}$). (b.) The wind rose at Pedra do Sal considering all events from September to November of 2013 (no filtered). Unique condition where winds blew only from 35.3$\,^{\circ}$ to 144.1$\,^{\circ}$ in three consecutive months.}\label{fig:windrose}
\end{figure}

The measurements were performed from September to November of 2013 in Pedra do Sal experiment located at flat coast in north-east Brazil (Fig. \ref{fig:map}). The site is chosen for wind energy purpose, so the instruments are installed next to a wind farm where 20 wind turbines with 55m hub height are aligned with the coast (Fig.\ref{fig:windrose}).

The wind flow is influenced by the trade winds and sea breeze \cite{oliver2005}. When the sea breeze intensity increases, the wind turns to north-east and when its decreases the wind turns to southeast. This diurnal cycle is very regular, so the wind direction varies just from $40^{\circ}$ to $120^{\circ}$ (Fig.\ref{fig:windrose}).

The sector $0\,^{\circ}$ to $90\,^{\circ}$ is defined according to International Electrotechnical Commission (IEC) recommendation \cite{IEC61400}, which selects the data that are not influenced by wind turbines wakes. The fetch in this sector comes from the ocean, where the roughness surface is dynamic and depends on the Reynolds stress \cite{charnock1955}. The roughness length ($z_{o}$) is estimated by fitting the logarithmic equation (Eq.~\ref{eq:prof}) on averaged observed profiles \cite{garratt1992}. The air flow also cross the shoreline, and an internal boundary layer (IBL) is develop at low levels due differences of roughness surface and sensible heat fluxes of sea-land interface. Considering the lidar position, the depth of the IBL is estimated to be about 40~m as suggested the height-fetch ratio of about 1/10 in convective condition \cite{garratt1990}.

\subsection{Measurements}

Pedra do Sal experiment has one meteorological tower of 100m height with five levels measurements and one medium range lidar wind profiler, see details on Table \ref{tab:sensors}. The lidar is positioned about 100m in front of the turbines (upwind), and 565m away from the tower (Fig \ref{fig:windrose}). The mounting of instruments on the tower are in accordance to IEC norm \cite{IEC61400}, and the cup anemometers and the ultrasonic are calibrated in wind tunnel.

\begin{center}
\begin{table}[ht]
\caption[]{Meteorological instruments installed at Pedra do Sal experiment. All sensors are mounted on the tower with 100~m height, except the lidar which is 565~m away from the tower.}\label{tab:sensors}
\centering
\small
\begin{tabular}{llrrrr}
\toprule
\multicolumn{1}{c}{\it Instrument}  & {\it Measurement}  & {\it Height} & {\it Scan } & {\it Average}\\
                                    &                    & {\it  [m]}   & {\it   [s]} & {\it [min]} \\
\midrule
Cup anemometer                      & Wind Speed                            & 10,40,60,80,98           & 1      & 10\\
Wind Vane                           & Wind Direction                        & 96 and 58                & 1      & 10\\
Thermohydrometer                    & Air Temperature and Humidity          & 98 and 40                & 1      & 10 \\
Barometer                           & Atmospheric Pressure                  & 13                       & 1      & 10\\
Sonic                               & Wind Speed  u,v,w \& Ts               & 100                      & 0.05   & 60 \\
Lidar                               & Wind Speed and Direction              & 26 $levels^{*}$          & 6      & 10 \\
\bottomrule
\end{tabular}\\
$^*$ Lidar first levels: 40,45,50,60. The next levels are every 20m from 80 to 500m.
\end{table}
\end{center}

\subsubsection{Lidar}

A pulsed lidar wind profile system, model WindCube8, is used to measure wind speed and direction profile from 40m to 500m height, see Table \ref{tab:sensors}. The lidar system calculates the horizontal and vertical components of the wind speed based on velocity azimuth display (VAD) technique. Each scan takes approximately 6~s to complete $360^{\circ}$ conical scanning, and the data are averaged and storage intervals of 10-min. The quality of the measurements are directly related to the signal of the carrier-to-noise ratio (CNR). When the atmosphere has poor aerosols content the signal is low and the data are considered suspected. This lidar is setup to discard measurements that has signal below −28~dB, and assures that only reliable measurements are saved. The data availability below 100\% in the 10-min interval are also storage, but only data with 100\% available is used. Data from 400~m to 500~m height is totally discard because of the poor signal and low data available. The system is level with its build in inclinometer, and a concrete basement is build due sand surface at local site. The lidar is aligned to true north with the gnomon at noon.

\subsubsection{Sonic}

Turbulence fluxes of momentum and heat are measured with 3D ultrasonic anemometer mounted at 100m height on the tower. The system has a datalogger that storage the vertical and horizontal components of wind and sonic temperature at 20 Hz. The sonic temperature is approximated to potential virtual temperature $ \theta_v$. The gas analyser of water vapor is not installed in the experiment, so the measurements are limited to a dry atmosphere. A Global Positioning System (GPS) is also used to synchronize the datalogger clock. The sensible heat flux and friction velocity are calculated using EddyPro software version 4.1 \cite{licor2012}. A linear detrending method is used to separate the eddy fluxes from the mean flow, and the planar fit method \cite{dijk2004} is used for tilt correction. The sonic positioned at 100m is chosen to avoid the influence of the IBL and flow distortion caused by the tower. The period of 60-min is used to calculate the fluxes for measurements above 100m height \cite{lee2004}.

\subsection{Dataset and filter}

Three data are organized in three dataset. The first file is the lidar data where 26 levels of wind speed and direction are storage in intervals of 10-min average. The second dataset contains parameters from the meteorological tower. The measurements are scan every one second by the datalogger and processed in intervals of 10-min. The third dateset cames from the Eddypro software that calculates the fluxes in intervals of 60-min. The fluxes are linear interpolated to intervals of 10-min, so the three dataset are formatted at the same time interval.

An algorithm is developed to filter the suspected data and it consists of three steps. The first step is to select the sector that is free of wake and obstacles which is from $0\,^{\circ}$ to $90\,^{\circ}$ as mention before. The second filter is to check lidar data that has 100\% available from 40m to 400m. If one level has data available lower than 100\%, the entire profile is discarded in this interval. The last filter is the quality control (QC) used in the Eddypro software \cite{mauder2006}. It is a complete QC that checks the limits of the data measured, spikes, time series statistics, and tests of the stead state condition and integral turbulence characteristics \cite{lee2004}. The flag 0 is excelent, 1 is satisfactory and 2 is bad \cite{licor2012}. The third filter eliminates any data with flag 2 from sensible heat flux or friction velocity flags. The algorithm  all parameters of the three dataset at this interval is considered invalid.

\subsubsection{Method of bins}

The method of bins is used to calculate the average and standard deviation of the wind profile and reference shear. The wind profile considers the wind speed at 10~m taken from the cup anemometer mounted on tower and the wind speed from 40~m to 400~m height from the lidar. The wind speed at those levels are averaged into bins of  $u_{*}$ and $H_{s}$  centred on multiples of 0.5~$m~s^{-1}$ and 2~$W~m^{-2}$ respectively. The calculated wind profile is estimated by Eq.~\ref{eq:prof} and based on the average bins of heat flux and friction velocities. The reference shear is measured from the lidar with data of 10-min wind speed average at 40, 45, 50, 60, 80, 100~m height and approximated by Eq.\ref{eq:refshearR}. The geometric mean height is 63~m according to Eq.~\ref{eq:zp}. The calculated reference shear is estimated by Eq.~\ref{eq:refshearS} and based on the average bins of heat flux and friction velocities. The average sonic virtual temperature is $302.5~K$ with standard deviation of $0.3~K$, and the air density is $1.163~kg~m^{-3}$ with standard deviation of $0.003~kg~m^{-3}$. The uncertainty is evaluated by standard uncertainty (Type A) estimated  by the standards deviation divide by square root of number of event of each bin\cite{isogum2008}.

\subsubsection{Thermal Stability Class}

The stability is analyzed at different friction velocities into bins of $u_{*}$ at 0.2, 0.3 and 0.4$~m~s^{-1}$. Therefore, the wind profiles are separated into three range of heat flux which defines the thermal stability class in this study. It is considered the neutral stratified condition when heat flux is zero. Then a narrow range of heat flux is defined to be the neutral class from -3 to 4~$W~m^{-2}$. The positive values of heat flux greater than 4~$W~m^{-2}$ are considered thermally unstable, and negative values lower than -3~$W~m^{-2}$ are considered thermally stable.

\section{Results and Discussion}

\subsection{Data Analysis}

\subsubsection{Data Available}

A total of 13,104 intervals of 10-min is the complete data series of the experiment. The Table \ref{tab:filter} shows the number of the samples found for each filter criteria and the percentage that is considered invalid. After apply the filter, the dataset available is reduced to 3,296 intervals of 10-min, or 549~h and 20~min, which represents 25,16~\% of the period.

\begin{center}
\begin{table}[ht]
\centering
\small
\caption[]{Number of 10-min mean data discarded for each filter procedure from September to November of 2013 in Pedra do Sal experiment. The three filters are combined which result the total data rejected.}\label{tab:filter}
\begin{tabular}{lrrr}
\toprule
\multicolumn{1}{c}{\it Filter Procedure}  & {\it Number of Samples} & {\it \%}\\
\midrule
Sector free of wake           & 5,580 & 42.6      \\
Lidar CNR                     & 8,994 & 68.6       \\
EddyPro QC Flag               & 5,340 & 40.7       \\
Combined Filters              & 9,808 & 74.8       \\
\bottomrule
\end{tabular}
\end{table}
\end{center}

Some samples can be detected by more than one filter at same time. It happens when the wind cames from the south-east from 0~h to 10~h in the morning. At this direction, the flow is affected by the turbine wakes (sector filter) and the inhomogeneous conditions of the coast (EddyPro filter). In the early morning, there is significant variation of wind speed and temperature because the transition of wind direction from sea-land to land-sea, and conversely. This no-steady state condition is detected by the EddyPro QC software and filtered. The largest amount of data filtered in this period is responsible to the poor CNR lidar signal caused by poor aerosols content in the atmosphere and the accumulation of sand on the lidar lens.

\begin{figure}[ht]
\centering
\includegraphics[width=9cm]{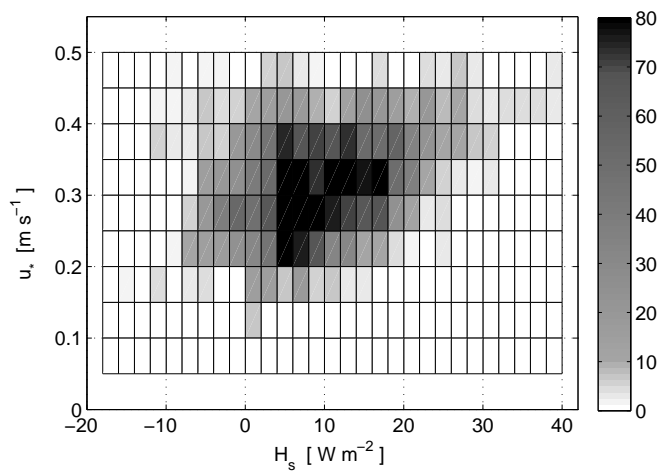}
\caption{Data available after filtering. The total data of 13,104 intervals of 10-min is reduced to 3,296 events. The gray scale indicates the number of events occurred for each bins of $H_s$ and $u_{*}$ (corrected). It does not represent the predominant stability condition as 74.84~\% of the events is filtered, but the short $H_s$ range and high $u_{*}$ indicate to a typical marine condition.} \label{fig:avalib}
\end{figure}

The data available in this experiment are separated in bins of sensible heat flux and friction velocity. The Fig.\ref{fig:avalib} shows the number of the events for each bin that is accounted in interval of 10~min. It does not represents the total distribution of the atmospheric conditions, because the filter applied. However, it is possible to observe that Pedra do Sal has a narrow range of heat flux because it is strongly influenced by the ocean.

\subsubsection{Wind Speed uncertainty}

\begin{figure}[ht]
\centerline{%
\begin{tabular}{c@{\hspace{0.01pc}}c}
\includegraphics[width=7.5cm]{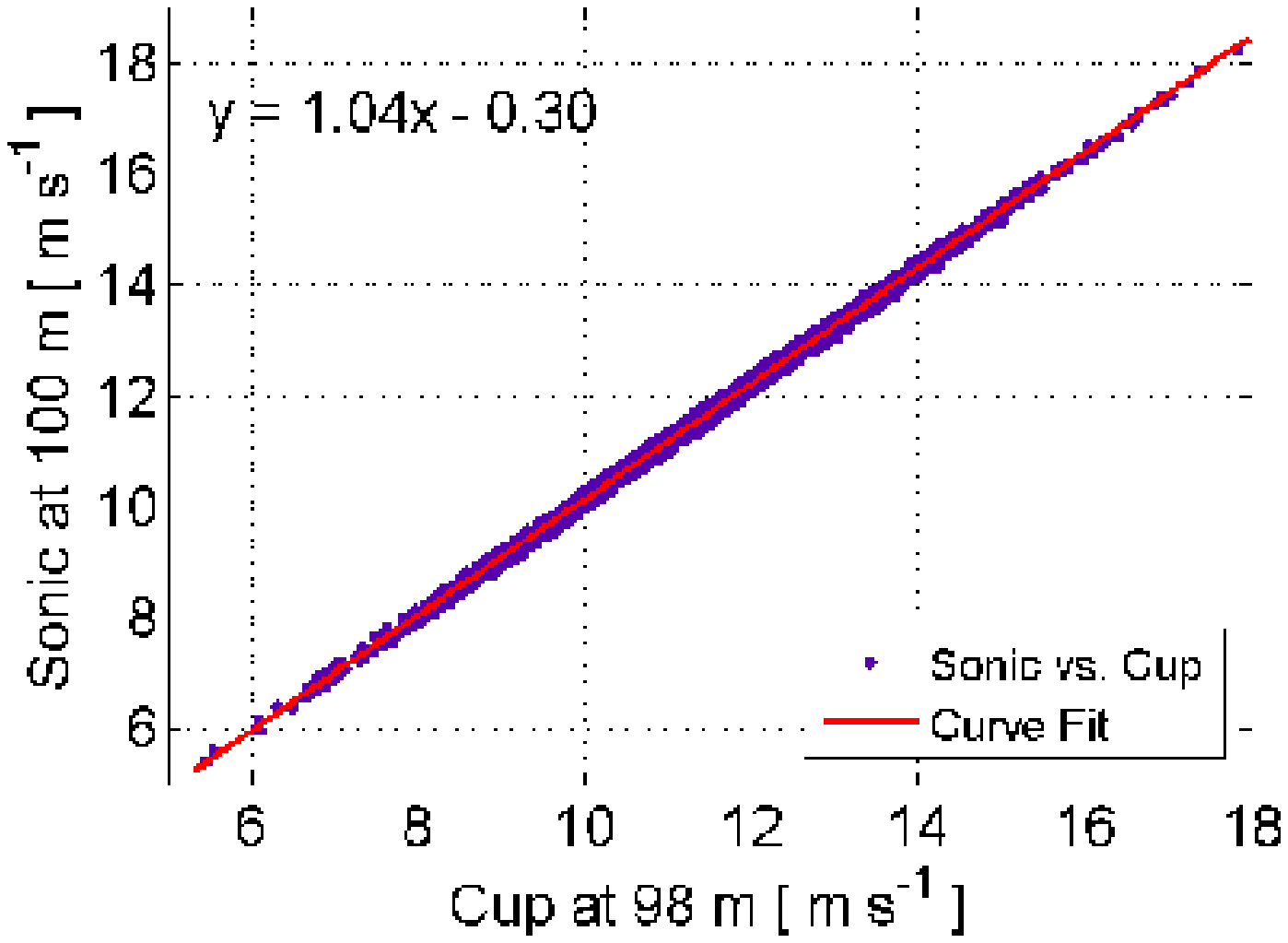} &
\includegraphics[width=7.5cm]{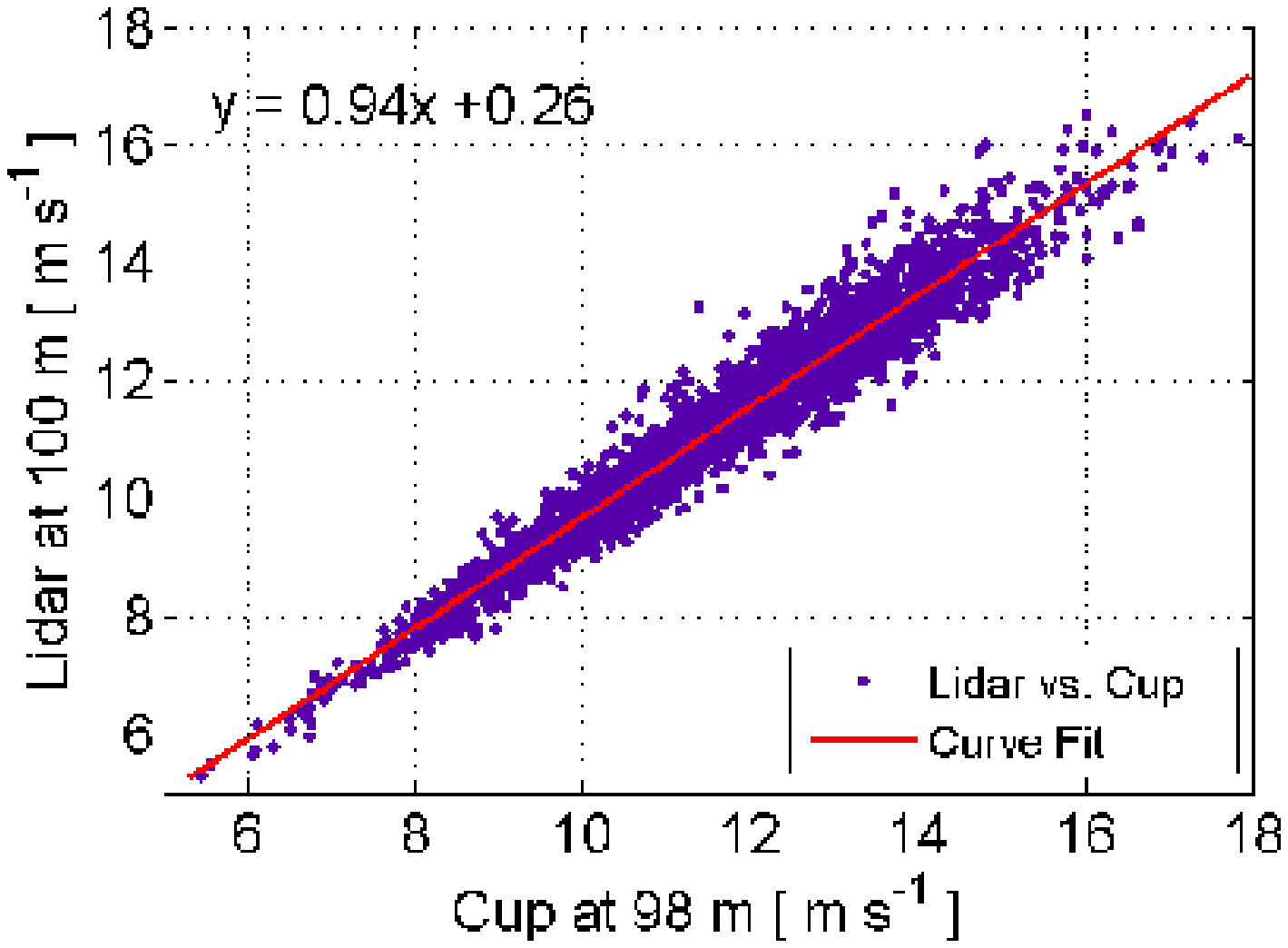} \\
a.  &  b.
\end{tabular}}
\caption{Comparison of wind speed instruments: (a.) sonic 3d at 100~m and cup anemometer at 98~m height, both mounted on the tower. (b.) lidar profiler at 100~m height installed 565~m away from the tower and cup anemometer at 98~m. The blue dots are averaged measurements taken into 10-min intervals and the red line is the linear fit}\label{fig:speed}
\end{figure}

The wind speed of lidar and sonic at 100m height are compared with calibrated cup anemomenter at 98m. The sonic and the cup anemometer are mounted on the same tower, and the lidar is 565m away from the tower. The sonic has an excellent coefficient of determination $R^{2}$ of 0.99 and root mean square error (RMSE) of 0.08~$m~s^{-1}$. Fig.\ref{fig:speed}~(a) confirms that the measurements are not affected by the structure of the tower. The slope is 4~\% higher is probably caused by the 2~m height difference. Fig.\ref{fig:speed} (b) shows the correlation of lidar and cup anemometer with $R^{2}$ of 0.95 and RMSE of 0.40~$m~s^{-1}$. The distance from lidar to tower can be the reason of the large dispersion and the slope of 6\% lower.

\subsubsection{von K\'{a}rm\'{a}n Constant}

\begin{figure}[ht]
\centerline{%
\begin{tabular}{c@{\hspace{0.01pc}}c}
\includegraphics[width=7.5cm]{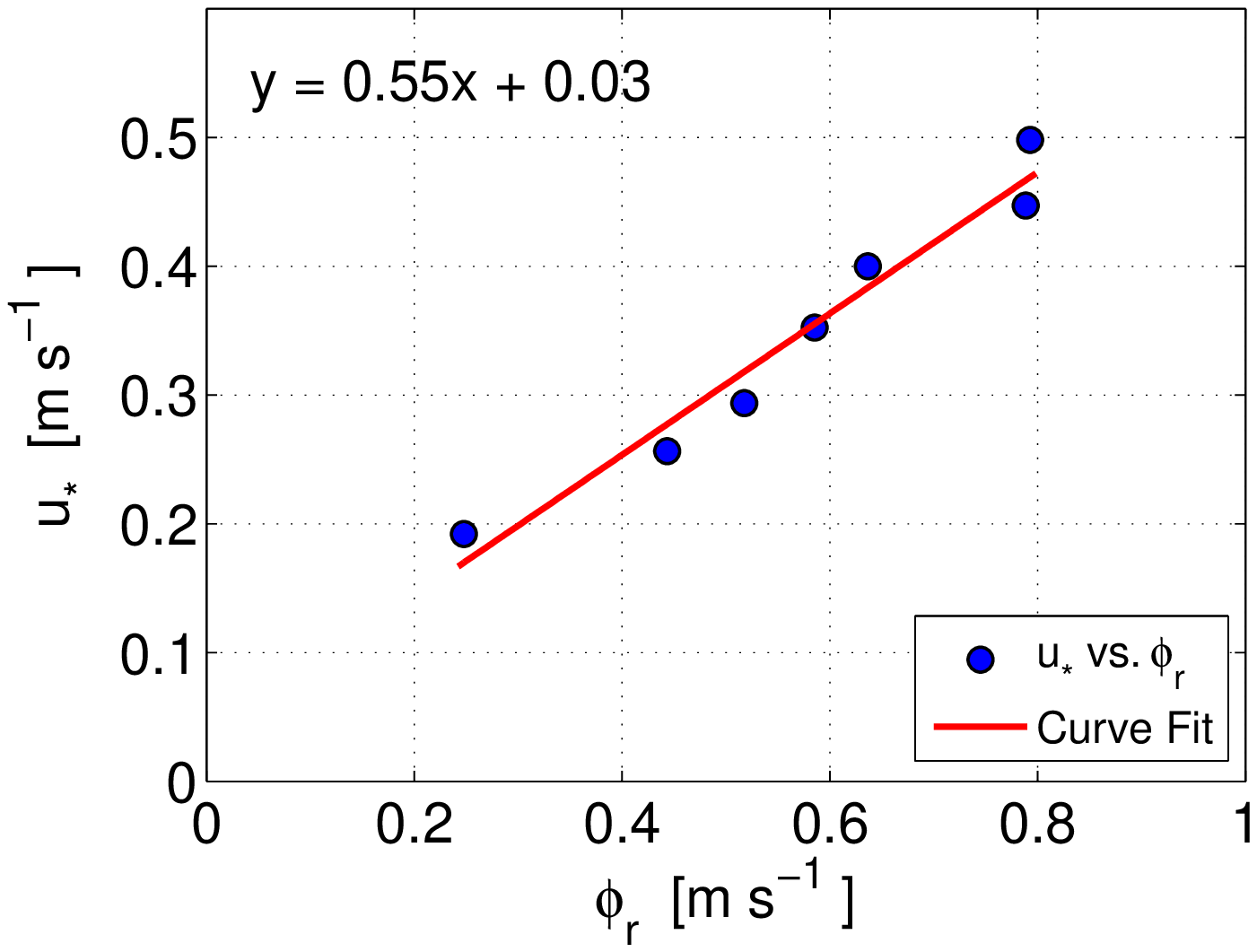} &
\includegraphics[width=7.5cm]{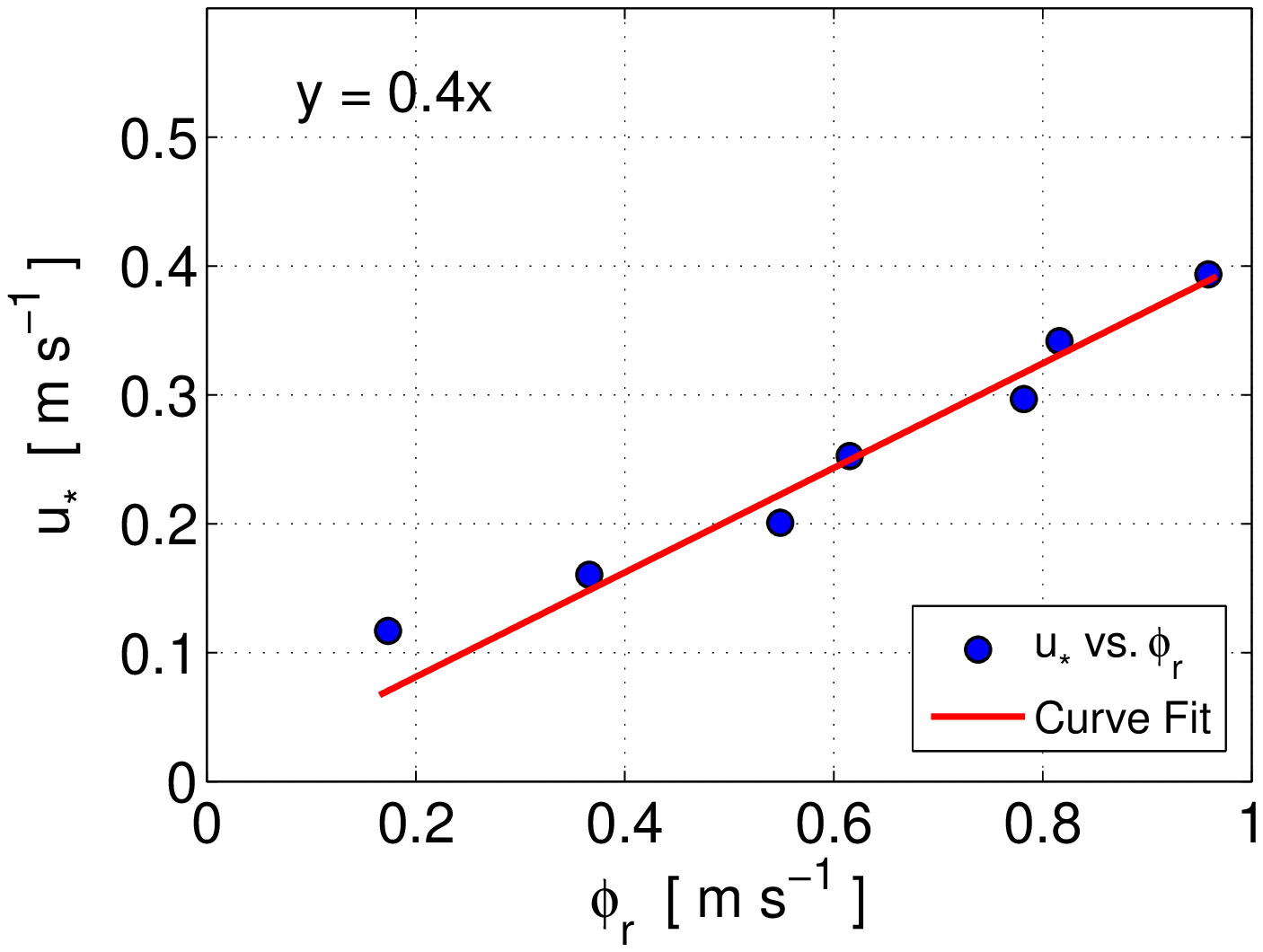} \\
a.~~ Uncorrected $u_{*}$ & b.~~ Corrected $u_{*}$
\end{tabular}}
\caption{The adjustment of $u_{*}$ based on von K\'{a}rm\'{a}n constant of 0.4. The blue dots indicate the averaged of $u_{*}$ and $\phi_{r}$ at neutral condition where $H_s$ range from -2~$W~m^{-2}$ to 2~$W~m^{-2}$. The red line is the linear fit between $u_{*}$ and $\phi_{r}$ ($z\partial{\bar{U}}/\partial{z}$) and the slope is the von K\'{a}rm\'{a}n constant. (a) Correlation of $u_{*}$ and $\phi_{r}$ without correction. (a) Correlation of $u_{*}$ and $\phi_{r}$ corrected with a factor of 0.66.   }\label{fig:karman}
\end{figure}

The preview analysis verify the von K\'{a}rm\'{a}n constant throughout the relation of reference shear and friction velocity at neutral condition. It is found 182 profiles between -2~$W~m^{-2}$ to 2~$W~m^{-2}$. The average sensible heat flux in this interval is 0.1~$W~m^{-2}$ and standards deviation of 0.4~$W~m^{-2}$. Fitting the curve by linear regression, it is found a slope of 0.55 and offset of 0.03, with $R^{2}$ of 0.96 and RMSE of 0.02~$m~s^{-1}$ (Fig.\ref{fig:karman}a.).

The boundary layer experiments such as Kansas \cite{kaimal1990} and Wangara \cite{Hicks1981}, also observed significant differences of von K\'{a}rm\'{a}n constant, and an large discussion is still open about the value of the constant and the measurements error related to those experiments. The von K\'{a}rm\'{a}n constant of 0.40 is the most value accepted today, and then errors are probably associated to the turbulent fluxes of momentum and heat \cite{foken2006}. In order to adjust the von K\'{a}rm\'{a}n to 0.4, the friction velocity need to be multiply by a factor of 0.66, which implies a difference of 34\% on the friction velocity. The slope is adjusted to a von K\'{a}rm\'{a}n of 0.40 with $R^{2}$ of 0.94 and RMSE of 0.02~$m~s^{-1}$ (Fig.\ref{fig:karman}b.). This adjustment also ensure that reference shear is equal to 1 when friction velocity reaches 0.4 which define the neutral condition.

\subsection{Wind Shear}

\begin{figure}[ht]
\centerline{%
\begin{tabular}{c@{\hspace{0.01pc}}c}
\includegraphics[width=7.5cm]{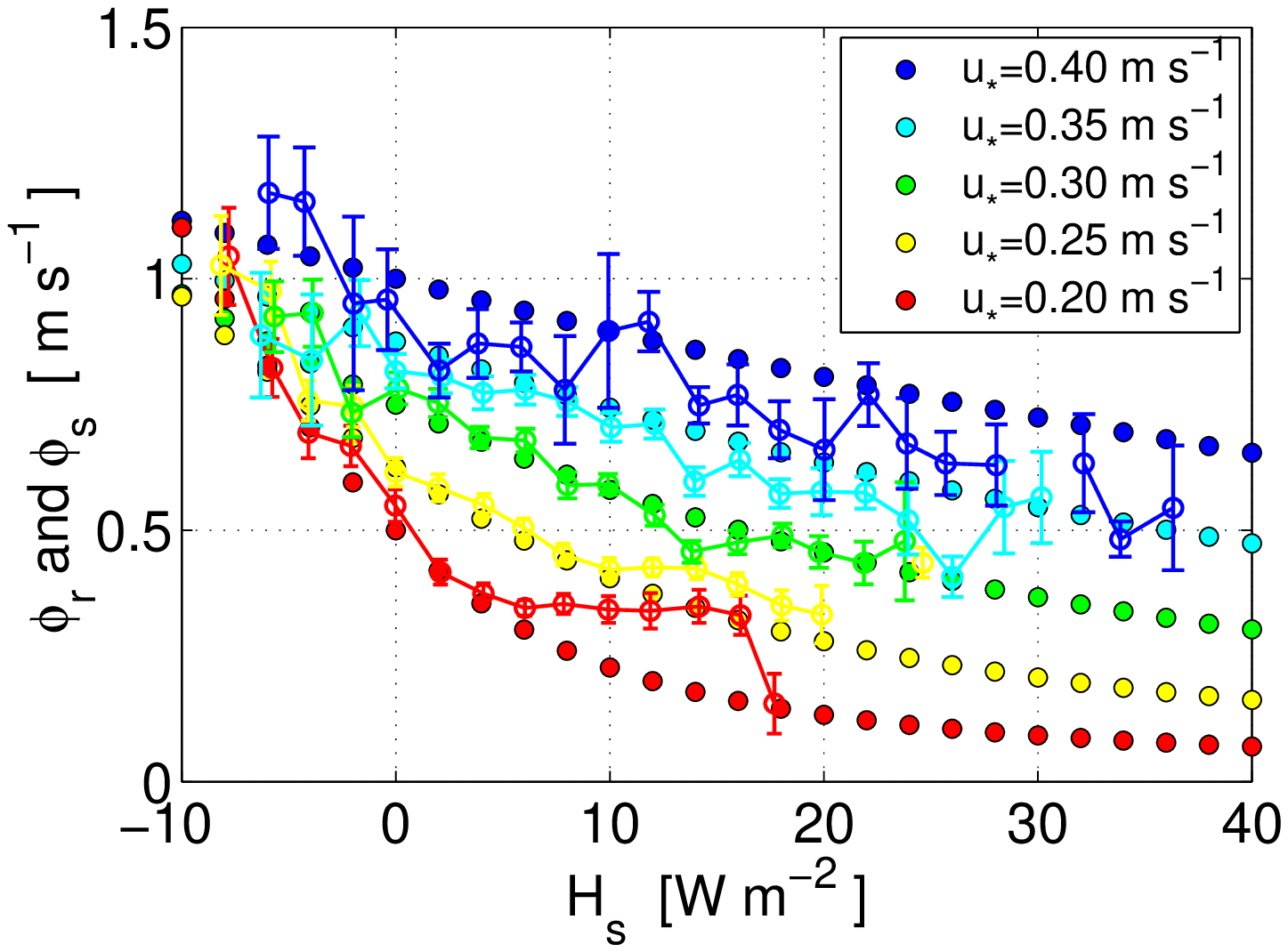} &
\includegraphics[width=7.5cm]{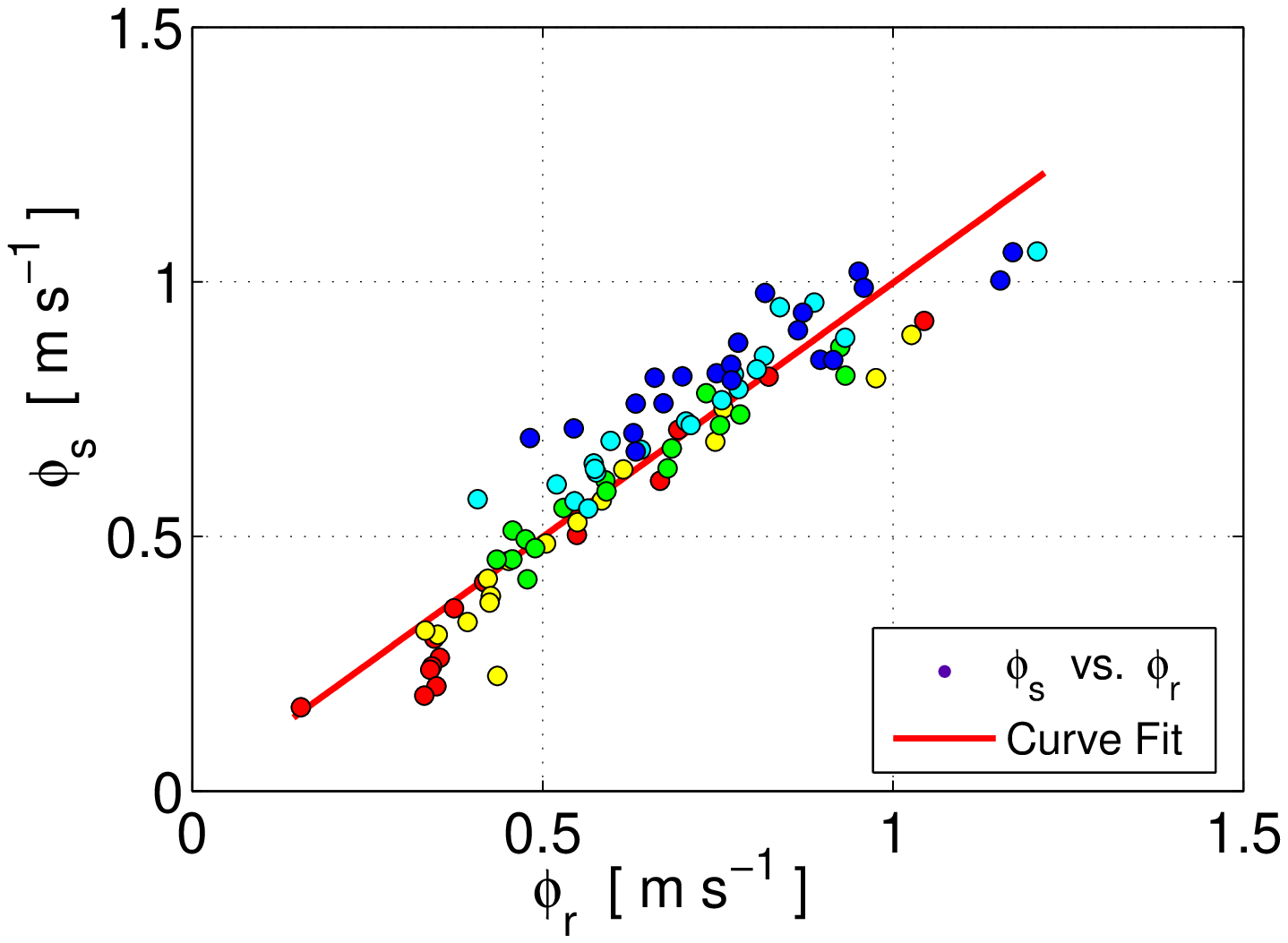} \\
a.~~ Reference Wind Shear & b.~~Correlation of Reference Shear
\end{tabular}}
\caption{ (a.) Comparison of reference shear observed and calculated at different $u_{*}$. The lines and empty circles represent the average $\phi_{r}$ and error bars indicate its standard uncertainty. The filled circles represent the $\phi_{s}$ calculated by the Eq. \ref{eq:refshearS}. (b.) The curve fit of observed and calculated reference shear adjusted at $Ri_s=1$} \label{fig:shear}
\end{figure}

Fig.~\ref{fig:shear}. shows the comparison of observed and calculated reference shears that are averaged into bins of sensible heat flux at different friction velocities. Some bins are discarded because of low number of samples available into the bin. The bins with less than 3 samples are not considered.

In the Fig.~\ref{fig:shear}a., it is observed that $\phi_{r}$ vs. $H_{s}$ relation depends on $u_{*}$. When $u_{*}$ is about 0.4~$m~s^{-1}$ it tends to be a linear curve. The large values of $u_{*}$ are usually associated with strong winds when the advection of momentum is dominated. In this case, the slope of wind profile is already oblique by the mechanical wind shear term and the influence of stability wind shear term is small. When $u_{*}$ is low, the reference shear turns to a exponential curve, because $u_{*}$ is inversely square in the stability wind shear term  as suggested in Eq.~\ref{eq:shear_total}. Therefore, wind profile is strongly influenced by $H_{s}$ at light winds and close to neutral condition. The $\phi_{r}$ curves converge to a upper limit about 1 when the heat flux is negative. An asymptotic $\phi_{r}$ curve is verified at unstable condition for all $u_{*}$ which means that wind profile lean to vertical slope ($\phi_{r}$ approach to zero) when $H_s$ increases. Large uncertainties of $\phi_{r}$ are found at $u_{*}>0.40$ and at $H_{s}<0$, because of strong mechanical turbulence and the exponential variation of $\phi_{r}$ at stable condition respectively. When $u_{*}>0.40~m~s^{-1}$, $\phi_{s}$ is overestimated and when $u_{*}<0.20~m~s^{-1}$ at unstable conditions $\phi_{s}$ is underestimated. The differences maybe caused by other terms of TKE equation that are not considered in the Eq.~\ref{eq:tke}. The parametrization of $Ri_s$ is estimate using linear regression by fitting $\phi_{s}$ vs. $\phi_{r}$ curve  (Fig.~\ref{fig:shear}.b) through the origin. The fitted curve has a slope of 1.00, with $R^{2}$ of 0.87 and RMSE of 0.08~$m~s^{-1}$ when $Ri_s$ is parameterized to a value of 1. It means that, in steady-state condition, the rate at which the energy is produced/consumed by the buoyant term is exactly equal to the rate at which the energy is produced/loss by the fraction of shear term affected by the thermal stratification.

\subsection{Wind Profile}

This section presents the comparison of theoretical and measured wind profile from 10m to 400m height at three different thermal stability condition. The Fig.~\ref{fig:prof1}, Fig.~\ref{fig:prof2}, Fig.~\ref{fig:prof3} show wind profiles separated into bins of friction velocities at 0.2, 0.3 and 0.4~$m~s^{-1}$ which are associated with light, moderated and strong wind speed respectively regarding to this site. The wind profile from the lidar is averaged for each stability class and error bars indicate the standard uncertainty. The calculated wind profile is based on averaged values of heat flux and friction velocity. The Tables \ref{tab:profile1}, \ref{tab:profile2} and \ref{tab:profile3} summarize the averaged values used in the Eq.~\ref{eq:prof}, and it also show the results of the correlation curve fit with observed wind profile. The roughness length ($z_{o}$) is estimated by fitting the calculated wind profile on averaged wind profiles at each thermal stability class. The maximum height of wind profile ($z_{max}$) is considered when it is still into the error bars, $1.01>slope>0.99$ and $R^{2}>0.87$ (same reference shear's $R^{2}$). The wind speed at 10~m is always different to the rest of the profile due the distance from the lidar and tower and it is also influenced by the internal boundary layer. Considering the underestimation of 6\% from Lidar vs.Cup anemometer comparison (Fig.~ \ref{fig:speed}b.), the highest wind speed difference of 5.9\% found between calculated and observed at 10~m is acceptable.

\subsubsection{Light Winds}

\begin{figure}[ht]
\centerline{%
\begin{tabular}{c@{\hspace{0.01pc}}c}
\includegraphics[width=7.5cm]{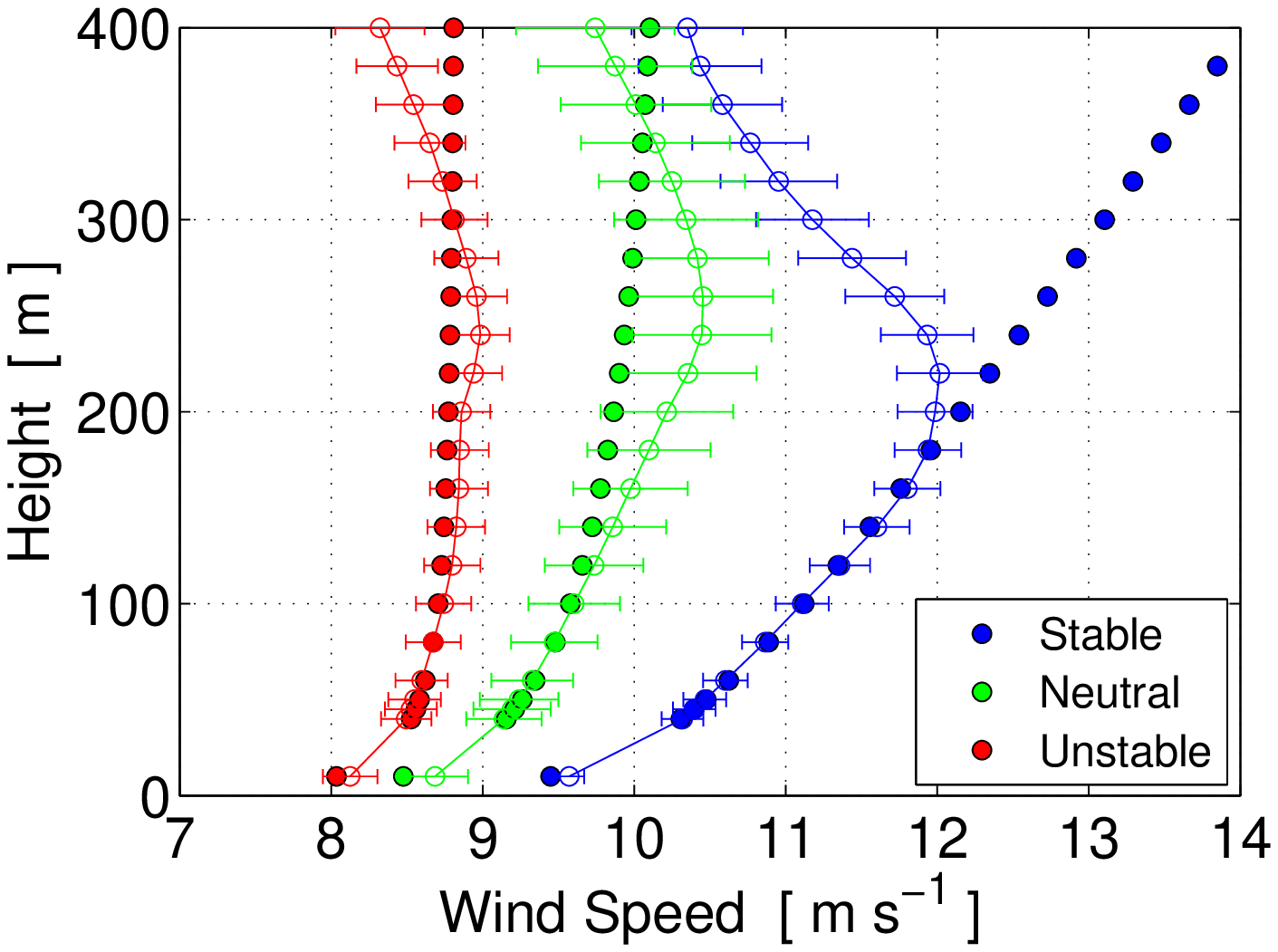} &
\includegraphics[width=7.5cm]{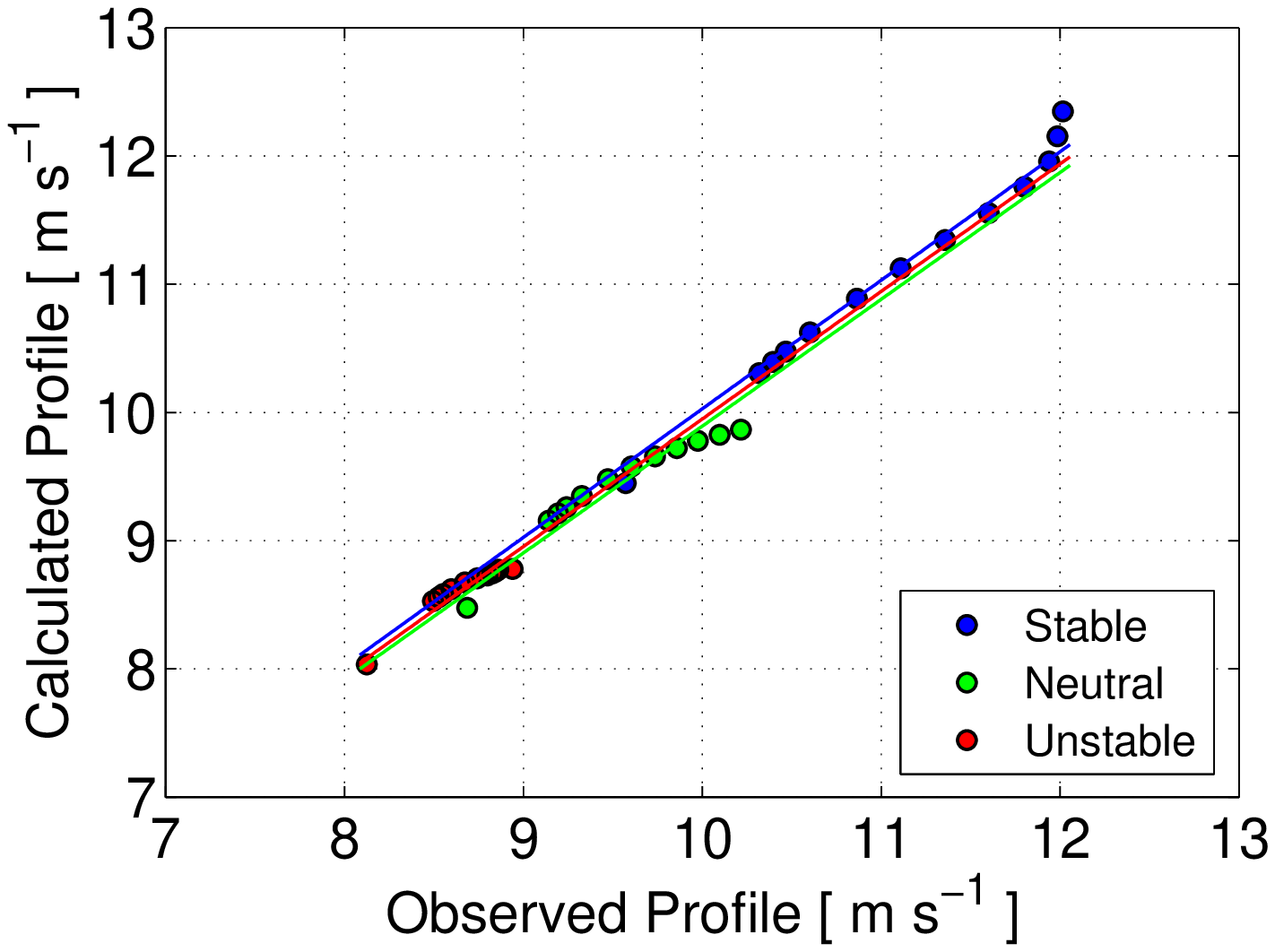} \\
a.~~Wind Profile at $u_{*}=0.20~m~s^{-1}$  & b.~~Correlation Eq.~\ref{eq:prof} vs. Lidar
\end{tabular}}
\caption{(a.) Comparison of observed and calculated wind profile from 10m to 400m height at three different thermal stability condition. The filled circles indicate the calculated wind profile based on averaged values of heat flux and friction velocity. The lines and empty circles indicate the average wind speed profile from the lidar and error bars represent its standard uncertainty of each stability class.(b.) The curve fit of observed and calculated wind profile for each stability condition. Each dot refers the wind speed at each respective height up to the maximum height.}\label{fig:prof1}
\end{figure}

Fig.~\ref{fig:prof1}a. shows that wind profile at light winds vary according to thermal stability class. The stratification of wind profile is evident at stable condition, and the theoretical wind profile shows a good correlation up to 220~m height. Above 220~m height, the wind speed decreases with height that is probably associated with the recirculation of the sea breeze.

At neutral condition, the standard uncertainty about $\pm 0.60~m~s^{-1}$ is large because of the rapid variation of the stability wind shear term when $H_s$ is close to zero. Thus, the neutral profile at light winds is better represented as an average transition from stable to unstable condition, but it still has a reasonable correlation up to 200~m height. At unstable condition, the wind profile has a slope almost vertical with wind speed difference of $0.37~m~s^{-1}$ from 40~m to 200~m height. At stable condition, this variation is $1.66~m~s^{-1}$.

\begin{center}
\begin{table}[ht]
\centering
\small
\caption[]{Summary of the averaged values used to calculate the wind profile at $u_{*} = 0.20~m~s^{-1}$, and the results of the correlation curve fit with observed wind profile.}\label{tab:profile1}
\begin{tabular}{cccccccc}
\toprule
Stability &   $u_{*}$    &    $H_{s}$     & $z_o$    & Slope & $R^{2}$  & RMSE        & $z_{max}$\\
 Class     & $[m~s^{-1}]$  & $[W~m^{-2}]$    & [m]    &  [-]  &  [-]     & [$m~s^{-1}$] & [m]     \\
\midrule
Stable   & $0.204  \pm 0.002$  & $-7 \pm 1$   & $0.10 \times 10^{-6}$  & 1.00 & 0.98  & 0.11 & 220\\
Neutral  & $0.202  \pm 0.002$  & $1  \pm 1$   & $0.50 \times 10^{-6}$  & 0.99 & 0.89  & 0.13 & 200\\
Unstable & $0.208  \pm 0.002$  & $15 \pm 2$   & $1.60 \times 10^{-6}$  & 0.99 & 0.90  & 0.06 & 220\\
\bottomrule
\end{tabular}\\
\end{table}
\end{center}

\subsubsection{Moderate Winds}

\begin{figure}[ht]
\centerline{%
\begin{tabular}{c@{\hspace{0.01pc}}c}
\includegraphics[width=7.5cm]{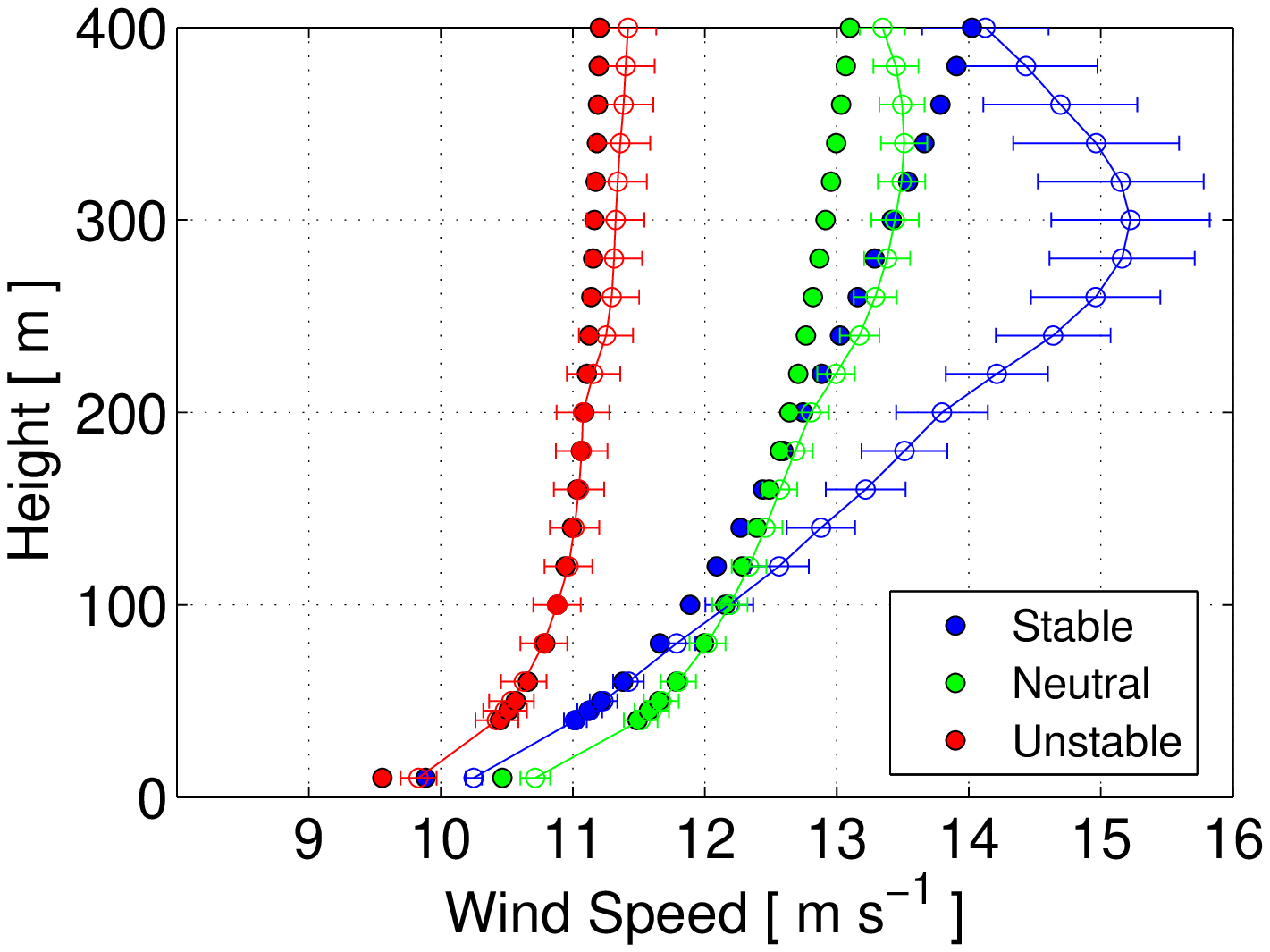} &
\includegraphics[width=7.5cm]{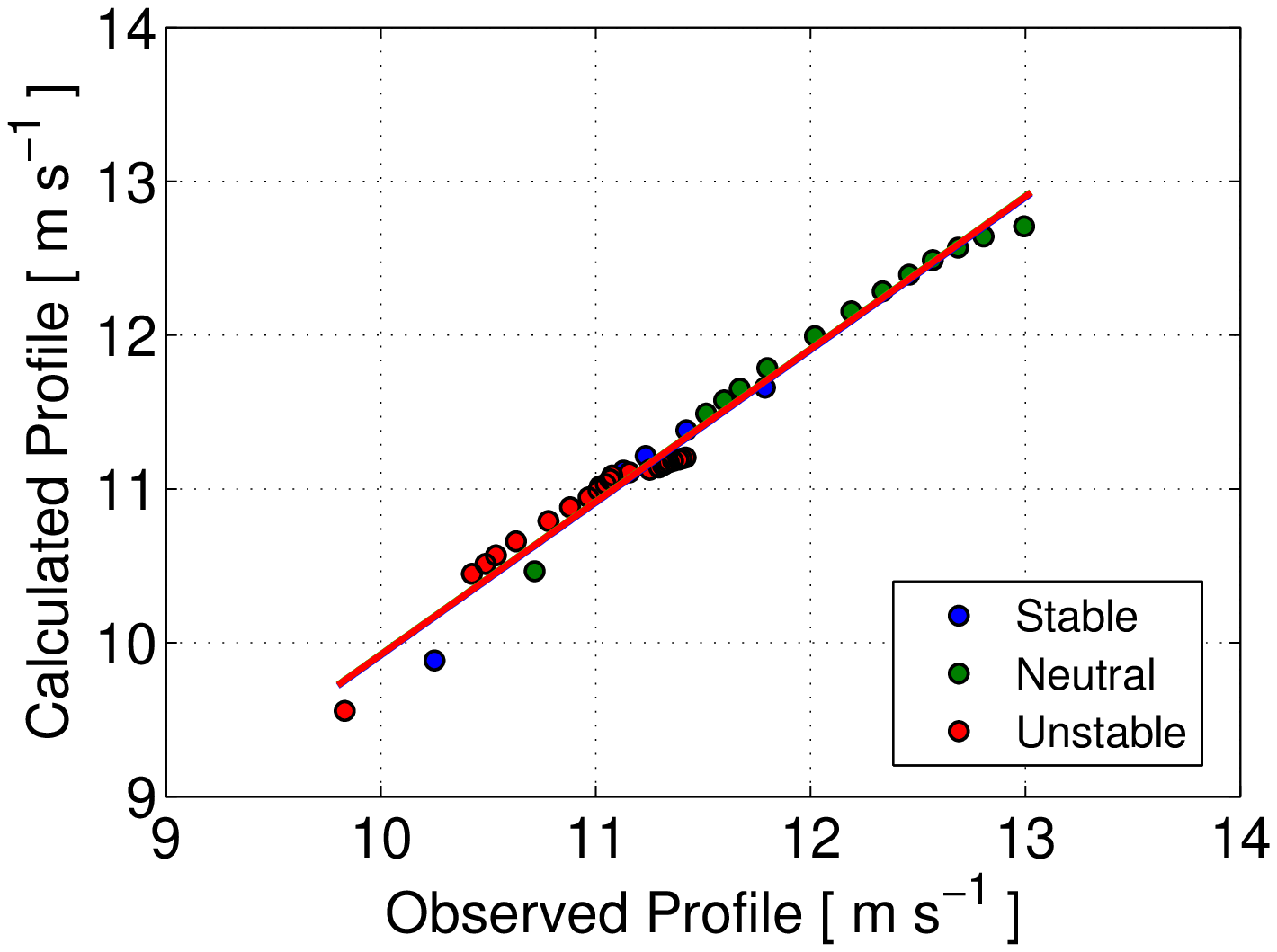} \\
a.~~Wind Profile at $u_{*}=0.30~m~s^{-1}$  & b.~~Correlation Eq.~\ref{eq:prof} vs. Lidar
\end{tabular}}
\caption{ (a.) Comparison of observed and calculated wind profile from 10m to 400m height at three different thermal stability condition. The filled circles indicate the calculated wind profile based on averaged values of heat flux and friction velocity. The lines and empty circles indicate the average wind speed profile from the lidar and error bars represent its standard uncertainty of each stability class.(b.) The curve fit of observed and calculated wind profile for each stability condition. Each dot refers the wind speed at each respective height up to the maximum height.}\label{fig:prof2}
\end{figure}

In Fig.~\ref{fig:prof2}, the correlation of wind profile at unstable and neutral condition is good (Table~ \ref{tab:profile2}). However, at stable condition, the calculated wind profile does not have a good agreement with observed wind profile (Stable Curve in Fig.~\ref{fig:prof2}).

It is observed a very stable condition wind profile which is not common in an oceanic boundary layer. The analysis of this weather conditions shows a particular situation that combines the mesoscale flow deviated by the topograph in the region, a concave coast and a warm advection from continent to ocean that transport warm air over a colder sea surface (Fig.~\ref{fig:map}a.). It happens in the afternoon when the trade winds still prevail and sea breeze is not strong enough to turn the winds from ocean to continent. In this case, the sensible heat flux measured from the sonic maybe is not representative of this air transported from different meteorological condition. When the heat flux is adjusted to -25~$W~m^{-2}$ the calculated wind profile at stable condition has a better correlation up to 300~m.

\begin{center}
\begin{table}[ht]
\centering
\small
\caption[]{Summary of the averaged values used to calculate the wind profile at $u_{*} = 0.30~m~s^{-1}$, and the results of the correlation curve fit with observed wind profile.}\label{tab:profile2}
\begin{tabular}{cccccccc}
\toprule
Stability &   $u_{*}$    &    $H_{s}$     & $z_o$    & Slope    & $R^{2}$  & RMSE        & $z_{max}$  \\
Class       & $[m~s^{-1}]$  & $[W~m^{-2}]$    & [m]  &   [-]     & [-]      & [$m~s^{-1}$] & [m]   \\
\midrule
Stable   & $0.303  \pm 0.003$  & $-9 \pm 1$   & $2.2 \times 10^{-5}$  & 0.99  & 0.94  & 0.14 & 80    \\
Neutral  & $0.298  \pm 0.001$  & $1 \pm 1$    & $1.0 \times 10^{-5}$  & 0.99  & 0.98  & 0.09 & 220  \\
Unstable & $0.302  \pm 0.002$  & $19 \pm 2$   & $4.5 \times 10^{-5}$  & 0.99  & 0.93  & 0.10 & 400   \\
\bottomrule
\end{tabular}\\
\end{table}
\end{center}

\subsubsection{Strong Winds}

\begin{figure}[ht]
\centerline{%
\begin{tabular}{c@{\hspace{0.01pc}}c}
\includegraphics[width=7.5cm]{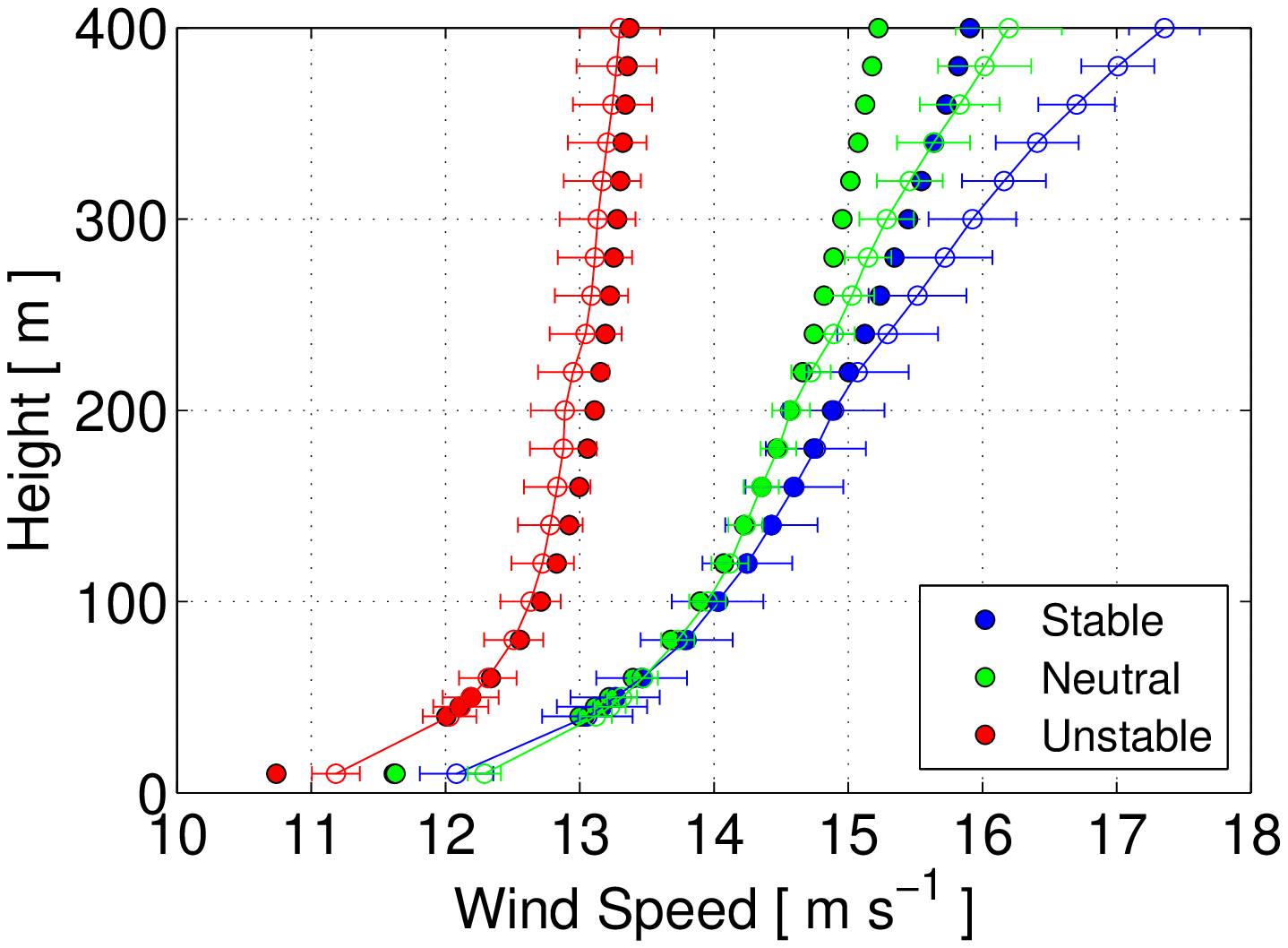} &
\includegraphics[width=7.5cm]{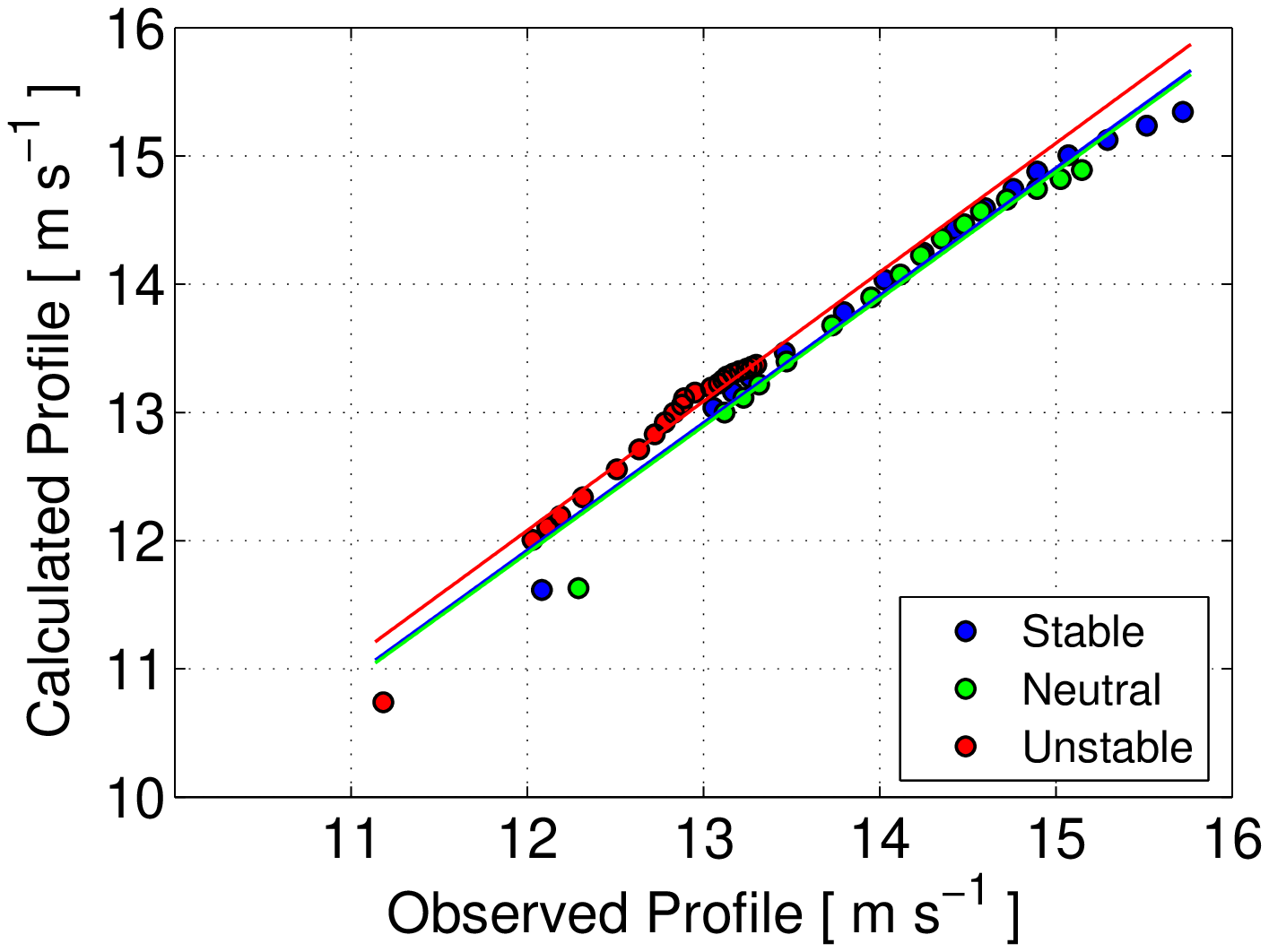} \\
a.~~Wind Profile at $u_{*}=0.40~m~s^{-1}$  & b.~~Correlation Eq.~\ref{eq:prof} vs. Lidar
\end{tabular}}
\caption{(a.) Comparison of observed and calculated wind profile from 10m to 400m height at three different thermal stability condition. The filled circles indicate the calculated wind profile based on averaged values of heat flux and friction velocity. The lines and empty circles indicate the average wind speed profile from the lidar and error bars represent its standard uncertainty of each stability class. (b.) The curve fit of observed and calculated wind profile for each stability condition. Each dot refers the wind speed at each respective height up to the maximum height.}\label{fig:prof3}
\end{figure}

Fig.~\ref{fig:prof3} shows the excellent correlation between the calculated and observed wind profile at strong winds in all thermal stability class. The advection of momentum is dominant and the influence of stability wind shear term is small because $u_{*}$ is high. Thus wind profile is already oblique due the mechanical shear term and the variation of $H_s$ does not significant affect the slope of the wind profile.

\begin{center}
\begin{table}[ht]
\centering
\small
\caption[]{Summary of the averaged values used to calculate the wind profile at $u_{*} = 0.40~m~s^{-1}$, and the results of the correlation curve fit with observed wind profile.}\label{tab:profile3}
\begin{tabular}{cccccccc}
\toprule
Stability &   $u_{*}$    &    $H_{s}$     & $z_o$   &  Slope & $R^{2}$  & RMSE        & $z_{max}$     \\
Class      & $[m~s^{-1}]$  & $[W~m^{-2}]$    & [m]  &  [-]  &  [-]     & [$m~s^{-1}$] & [m]   \\
\midrule
Stable   & $0.397  \pm 0.006$  & $-8  \pm 1$  & $0.85 \times 10^{-4}$  & 0.99 & 0.97  & 0.15 & 280  \\
Neutral  & $0.396  \pm 0.002$  & $1   \pm 1$  & $0.80 \times 10^{-4}$  & 0.99 & 0.96  & 0.16 & 280  \\
Unstable & $0.398  \pm 0.003$  & $23  \pm 2$  & $2.00 \times 10^{-4}$  & 1.01 & 0.96  & 0.13 & 400    \\
\bottomrule
\end{tabular}\\
\end{table}
\end{center}

The difference of the wind speed from 40~m to 200~m height is $0.85~m~s^{-1}$ at unstable condition, and $1.86~m~s^{-1}$ at stable condition. Above 220~m height, the wind speed increases with height faster than it is theoretic expect which is possible associated with the recirculation of the land breeze.

\section{Summary and Conclusion}

This paper proposes a stability wind shear term of logarithmic wind profile equation that model the wind profile according to the atmospheric stability. The theory is based on three new assumptions. First, the Reynolds stress can be divided in two distinct terms associated with wind shear of mechanical and thermal stratification process, considering eddy viscosity coefficient at neutral condition. Second, the mechanical wind shear is affected by the same term of thermal stratification process at no neutral condition. Third, the ratio between the buoyant terms of TKE equation and the fraction of the shear term affected by thermal stratification process is considered constant at steady state ($Ri_s$). The proposed equation performs well in comparison with the results from Pedra do Sal experiment regarding to corrections of friction velocity and uncertainties of the lidar. The derivative of wind profile shows an excellent correlation when analyzed by reference shear. The lack of data at very strong and very light winds difficult to validate the theory at extreme conditions, but the results show that at unstable conditions, the theoretical wind shear tend to be overestimated at strong winds and underestimated at light winds. The wind profile is well predicted in most of the meteorological conditions from light to strong winds and at stable, neutral and unstable conditions. An exception is observed at very stable condition when the fluxes measured at the site are probably not representative of the air transported from different meteorological condition. The theoretical wind profile agrees with observed wind profile up to 400~m at unstable condition and up to 280~m at stable condition. The results suggests that the top of surface layer can vary from 200~m to 400~m depends on the mesoscale circulation of sea and land breeze. The low influence of Coriolis force maybe contribute to a deeper surface layer (Prandtl layer) at this site. The particular meteorological conditions of this site limit the theory to be valid only to this specific site. The proposed logarithmic profile equation can collaborate wind energy industry to better estimate the wind profile considering the influence of atmospheric stability.

\ack This study is supported by the Brazilian Electricity Regulatory Agency (ANEEL) under R\&D project $0403-0020/2011$, funded by Tractebel Energia S.A. (GDF Suez). Authors also aknowledge support from the National Council for Scientific and Technological Development of Brazil (CNPq).


\bibliographystyle{wileyj}
\bibliography{Profile}

\end{document}